\documentclass[aps,prd,nofootinbib,11pt]{revtex4}
\usepackage{amsfonts}
\usepackage{txfonts}
\usepackage{graphicx}
\usepackage{dcolumn}
\usepackage{bm}
\usepackage{amssymb}
\usepackage{latexsym}

\newcommand{\be}{\begin{equation}}
\newcommand{\ee}{\end{equation}}
\newcommand{\bq}{\begin{eqnarray}}
\newcommand{\eq}{\end{eqnarray}}

\begin{document}

\title{Stable de Sitter critical points of the cosmology in quadratic gravitation
with torsion}

\author{Guo-Ying Qi}
\address{College of physics and electronics, Liaoning Normal University, Dalian,
116029, China,\\
Purple Mountain Observation, Academia Sinica, Nanjing, 210008, China }

\begin{abstract}
Homogeneous isotropic spatial flat cosmological models with two torsion
functions in vacuum are built and investigated in the framework of de Sitter
gauge theory of gravity. It is shown that by certain choices of parameters
of gravitational Lagrangian the cosmological equations have some exact
constant solutions that turn out to be stable de Sitter critical points of
dynamical systems and can explain observable acceleration of cosmological
expansion. The role of the space-time torsion provoking the acceleration of
cosmological expansion is shown.

\end{abstract}

\maketitle

\section{Introduction}

Recently there has been a burst of activity dealing with quadratic
gravitation. For example, the curvature-squared terms added to the usual
Einstein action with cosmological constant have played a role in two recent
investigations of four-dimensional gravity: in critical gravity [1], and in
a pure Weyl-squared action considered by Maldacena [2].

The critical gravity provides a consistent toy model for quantum gravity as
a useful simplified arena for studying some aspects of a potentially
renormalisable theory of massless spin-2 fields in four dimensions.

The conformal gravity theory has been advanced as a candidate alternative to
standard Einstein gravity. As a quantum theory the conformal theory is both
renormalizable and unitary, with unitarity being obtained because the theory
is a PT symmetric rather than a Hermitian theory. Because the variation of
the conformal action leads to fourth-order equations of motion, it had long
been thought that the theory would not be unitary. However, as has been
shown by Bender and Mannheim [3] that one can find a realization of the
theory that is unitary. Consequently, conformal gravity is to be regarded as
a bona fide quantum gravitational theory. The conformal gravity theory can
quite naturally handle some of the most troublesome problems in physics, the
quantum gravity problem, the vacuum energy problem, and the dark matter
problem. [4]

As a modified gravity theory quadratic gravitation has been used in
cosmology [5]. In order to explain observable acceleration of cosmological
expansion some authors introduce torsion terms in quadratic gravitation [6].
The quantum aspects of torsion theory and the possibility of the space-time
torsion to exist and to be detected have been discussed in [7].The\
astronomical observations show that our universe is probably an
asymptotically de Sitter (dS) one with a positive cosmological constant $%
\Lambda $ [8]. If a gravitational theory of Yang-Mills type is constructed
starting from de Sitter gauge invariance principle, its gravitational
Lagrangian naturally turns out to be the one of quadratic gravitation with
torsion as will be shown in this paper. Therefore, a investigation of
quadratic gravitation with torsion and its cosmological solutions expressed
by de Sitter critic points will be carried out. The field equations will be
derived. These equations are quite different from the equations obtained
from Riemannian geometry based quadratic Lagrangians when varied with
respect to the metric. Applying to the space flat FRW cosmology some de
Sitter critical point solutions will be obtained. The stability of them will
be analyzed.

The paper is organized as follows. In section II, starting from a Clifford
algebra $C\left( 3,1\right) $ the gravitational Lagrangian of a de Sitter
gauge theory is constructed, the Lagrange equations of gravitational fields
are derived. Applying them to a spatial flat universe the cosmological
equations are obtained in section III. The vacuum solutions of these
equations in two specific cases are presented in section IV. These two
models correspond to the conformal cosmology of Mannheim [4,9] and the
zero-energy gravity of Deser and Tekin [10], respectively. In contrast to
them, the tetrad and the spin connection are taken to be the basic field
variables and the torsion plies a important role here. In these specific
models the cosmological equations are written as some dynamical systems, the
real de Sitter critical points of them are obtained. Among these points, the
stable ones which turn out to be exact constant solutions and describe the
asymptotic behavior of the universe are found. In section V some concluding
remarks are given. In Appendixes the calculations for stability analysis are
presented.

\section{Lagrangian and field equations}

We begin with a brief introduction of a de Sitter gauge theory. In a
gravitational gauge theory coupled to matter sources involving Dirac fields
it is convenient to take Dirac matrices $\gamma _I$ and their commutators $%
\sigma _{IJ}=\frac 12\left[ \gamma _I,\gamma _J\right] $ as the basis of the
gauge algebra. In this case we are led to a de Sitter gauge theory. Let $%
\left\{ \gamma _I\right\} \;\left( I=0,1,2,3\right) $ be a basis of an inner
product space with signature $\left( -,+,+,+\right) $. A Clifford algebra $%
C\left( 3,1\right) $ can be constructed by introducing the condition
\begin{equation}
\gamma _I\gamma _J+\gamma _J\gamma _I=2\eta _{IJ}I.
\end{equation}
with $\eta _{IJ}=$diag$(-1;1;1;1)$. There is a 10-dimensional subspace of $%
C\left( 3,1\right) $ which is a Lie algebra with basis $\gamma _5\gamma _I$
and $\sigma _{IJ}=\frac 12\left[ \gamma _I,\gamma _J\right] $. This is the
Lie algebra of a de Sitter group. We can introduce a connection [11,12]
\begin{equation}
\omega =\Gamma +\frac 1l\gamma _5{\bf e},
\end{equation}
defined by
\begin{equation}
{\bf e}=e{}^I{}_\mu \gamma _I\otimes dx^\mu ,
\end{equation}
and
\[
\Gamma =\frac 14\Gamma ^{IJ}{}_\mu \sigma _{IJ}\otimes dx^\mu ,
\]
where $l$ denotes a constant with the dimension of length. The curvature of $%
\omega $ is
\begin{equation}
\Omega =d\omega +\frac 12\left[ \omega ,\omega \right] ={\bf R}+\frac 1l%
\gamma _5{\bf T}-\frac 1{l^2}{\bf V},
\end{equation}
where
\begin{eqnarray}
{\bf R} &=&d\Gamma +\frac 12\left[ \Gamma ,\Gamma \right] ,  \nonumber \\
{\bf T} &=&d{\bf e}+\left[ \Gamma ,{\bf e}\right] ,  \nonumber \\
{\bf V} &=&\frac 12\left[ {\bf e},{\bf e}\right] .
\end{eqnarray}
The Lorentz curvature ${\bf R}$, the torsion ${\bf T}$, and the cosmological
term ${\bf V}$ are given by, respectively,
\begin{eqnarray}
{\bf R} &=&\frac 18R^{IJ}{}_{\mu \nu }\sigma _{IJ}\otimes dx^\mu \wedge
dx^\nu ,  \nonumber \\
{\bf T} &=&\frac 12T^I{}_{\mu \nu }\sigma _{IJ}\otimes dx^\mu \wedge dx^\nu ,
\nonumber \\
{\bf V} &=&e{}^I{}_\mu e{}^J{}_\nu \sigma _{IJ}\otimes dx^\mu \wedge dx^\nu ,
\end{eqnarray}
with
\begin{equation}
R^{IJ}{}_{\mu \nu }=\partial _\mu \Gamma {}^{IJ}{}_\nu -\partial _\nu \Gamma
{}^{IJ}{}_\mu +\eta _{KL}\Gamma {}^{IK}{}_\mu \Gamma {}^{LJ}{}_\nu -\eta
_{KL}\Gamma {}^{IK}{}_\nu \Gamma {}^{LJ}{}_\mu ,
\end{equation}
and
\begin{equation}
T{}^I{}_{\mu \nu }=\partial _\mu e{}^I{}_\nu -\partial _\nu e{}^I{}_\mu
+\Gamma {}^I{}_{J\mu }e{}^J{}_\nu -\Gamma {}^I{}_{J\nu }e{}^J{}_\mu .
\end{equation}

Based on the local gauge invariance principle the gravitational Lagrangian
can be made up of a quadratic term of the curvature $\Omega $ and its Hodge
dual $*\Omega $:
\begin{equation}
{\cal L}=-\frac 18Tr\left( *\Omega \wedge \Omega \right) =\left( \frac 1{32}%
R_{\mu \nu }{}^{\rho \sigma }R^{\mu \nu }{}_{\rho \sigma }-\frac 14%
l^{-2}T{}^\mu {}_{\nu \rho }T{}_\mu {}^{\nu \rho }+\frac 12%
l^{-2}R-12l^{-4}\right) e,
\end{equation}
where
\begin{equation}
e=\det \left| e^I{}_\mu \right| .
\end{equation}

{\em \ }In four dimensional spacetime the Gauss-Bonnet term $\sqrt{-g}\left[
R_{\mu \nu \lambda \tau }R^{\mu \nu \lambda \tau }-4R_{\mu \nu }R^{\mu \nu
}+R^2\right] $ is purely topological and then the Lagrangian can be taken as
\begin{equation}
{\cal L}=-\frac 18Tr\left( *\Omega \wedge \Omega \right) =\left( \frac 18%
R_{\mu \nu }R^{\mu \nu }-\frac 1{32}R^2-\frac 14l^{-2}T{}^\mu {}_{\nu \rho
}T{}_\mu {}^{\nu \rho }+\frac 12l^{-2}R-12l^{-4}\right) e.
\end{equation}
For the sake of a neater argument we extend the Lagrangian to including the
coefficients
\begin{equation}
\beta =\frac 18l^2,\alpha =-\frac 1{32}l^2,\gamma =-\frac 14,
\end{equation}
and rewrite (10) as
\begin{equation}
{\cal L}=\left( \beta l^{-2}R_{\mu \nu }R^{\mu \nu }+\alpha l^{-2}R^2+\gamma
l^{-2}T{}^\mu {}_{\nu \rho }T{}_\mu {}^{\nu \rho }+\frac 12%
l^{-2}R-12l^{-4}\right) e=Le,
\end{equation}
with
\begin{equation}
L=\beta l^{-2}R_{\mu \nu }R^{\mu \nu }+\alpha l^{-2}R^2+\gamma l^{-2}T{}^\mu
{}_{\nu \rho }T{}_\mu {}^{\nu \rho }+\frac 12l^{-2}R-12l^{-4}.
\end{equation}
${\cal L}$ is just the Lagrangian of quadratic-curvature gravities [10] with
torsion.

The variational principle yields the field equations for the tetrad{\em \ } $%
e_I{}^\mu $ and the spin connection $\Gamma {}^{IJ}{}_\mu $

\begin{eqnarray}
\frac{\delta {\cal L}}{\delta e_I{}^\mu } &=&eE^I{}_\mu ,  \nonumber \\
\frac{\delta {\cal L}}{\delta \Gamma ^{IJ}{}_\mu } &=&es_{IJ}{}^\mu ,
\end{eqnarray}
where $E^I{}_\mu $ and $s_{IJ}{}^\mu $ are energy- momentum and spin tensors
of the matter source, respectively, the variational derivatives are given by
\begin{eqnarray}
&&\frac{\delta {\cal L}}{\delta e_I{}^\mu }  \nonumber \\
&=&\{\beta l^{-2}\left( 2e{}^{I\sigma }R{}^\rho {}_\sigma R{}_{\rho \mu
}+2e^J{}_\rho R{}^{\rho \sigma }{}R{}^I{}_{J\mu \sigma }-e{}^I{}_\mu R_{\rho
\sigma }R^{\rho \sigma }\right) +\alpha l^{-2}\left( 4e^{I\nu }R{}_{\nu \mu
}-e{}^I{}_\mu R\right) R  \nonumber \\
&&+\gamma l^{-2}\left( 4e{}^{I\nu }T{}^\lambda {}_{\nu \tau }T{}{}_{\lambda
\mu }{}^\tau -4\partial _\nu \left( e{}^{I\lambda }T{}_{\mu \lambda }{}^\nu
\right) -e{}^I{}_\mu T{}^\lambda {}_{\rho \sigma }T{}_\lambda {}^{\rho
\sigma }+\left( 4e{}^{I\lambda }T{}_{\mu \lambda }{}^\nu \right)
e{}^K{}_\tau \partial _\nu e_K{}^\tau \right)  \nonumber \\
&&+l^{-2}\left( e^{I\nu }R{}_{\nu \mu }-\frac 12e{}^I{}_\mu R\right)
+12l^{-4}e{}^I{}_\mu \}e,
\end{eqnarray}
\begin{eqnarray}
\frac{\delta {\cal L}}{\delta \Gamma ^{IJ}{}_\mu } &=&\{2\beta
l^{-2}e_J{}^\lambda [e_I{}^\mu \partial _\nu R_\lambda {}^\nu -e_I{}^\nu
\partial _\nu R_\lambda {}^\mu +\left( e_I{}^\nu R_\lambda {}^\mu -e_I{}^\mu
R_\lambda {}^\nu \right) e{}^K{}_\tau \partial _\nu e_K{}^\tau  \nonumber \\
&&+e_I{}^\tau \Gamma ^\nu {}_{\nu \tau }R_\lambda {}^\mu +e_I{}^\nu \Gamma
^\tau {}_{\nu \lambda }R_\tau {}^\mu -e_I{}^\mu \Gamma ^\tau {}_{\nu \lambda
}R_\tau {}^\nu -e_I{}^\tau \Gamma ^\mu {}_{\nu \tau }R_\lambda {}^\nu ]
\nonumber \\
&&+2\alpha l^{-2}[\left( e_I{}^\nu e_J{}^\tau -e_J{}^\nu e_I{}^\tau \right)
\Gamma ^\mu {}_{\nu \tau }R+\left( e_J{}^\mu e_I{}^\nu -e_I{}^\mu e_J{}^\nu
\right) \left( \Gamma ^\lambda {}_{\lambda \nu }R-\partial _\nu R\right)
\nonumber \\
&&+\left( e_I{}^\nu e_J{}^\mu -e_I{}^\mu e_J{}^\nu \right) Re{}^K{}_\tau
\partial _\nu e_K{}^\tau ]+4\gamma l^{-2}e_{I\nu }e{}_J{}^\tau T{}^{\nu \mu
}{}_\tau  \nonumber \\
&&+\frac 12l^{-2}[\left( e_I{}^\nu e_J{}^\tau -e_J{}^\nu e_I{}^\tau \right)
\Gamma ^\mu {}_{\nu \tau }+\left( e_I{}^\nu e_J{}^\mu -e_I{}^\mu e_J{}^\nu
\right) \left( \Gamma ^\lambda {}_{\lambda \nu }+e{}^K{}_\tau \partial _\nu
e_K{}^\tau \right) ]\}e.
\end{eqnarray}
That may be, the two main field equations are rather complicated. They
really look nothing like the familiar, well-analyzed equations of GR. To
help understand the significance of these new relations, and to use our
previous experience, we will do a translation of (16,17) into a certain
effective Riemannian form--transcribing from quantities expressed in terms
of the tetrad $e_I{}^\mu $ and spin connection $\Gamma {}^{IJ}{}_\mu $ into
the ones expressed in terms of the metric $g_{\mu \nu }$ and torsion $%
T^\lambda {}_{\mu \nu }$ (or contortion $K^\lambda {}_{\mu \nu }$).

As is well-known, the affine connection $\Gamma ^\lambda {}_{\mu \nu }$ can
be represented in the form
\begin{eqnarray}
\Gamma ^\lambda {}_{\mu \nu } &=&e_I{}^\lambda \partial _\mu e^I{}_\nu
+e_J{}^\lambda e^I{}_\nu \Gamma {}^J{}_{I\mu }  \nonumber \\
&=&\left\{ _\mu {}^\lambda {}_\nu \right\} +K^\lambda {}_{\mu \nu },
\end{eqnarray}
where $\left\{ _\mu {}^\lambda {}_\nu \right\} $, $K^\lambda {}_{\mu \nu }$
are the Christoffel symbol and the contortion, separately, with
\begin{eqnarray}
K^\lambda {}_{\mu \nu } &=&-\frac 12\left( T^\lambda {}_{\mu \nu }+T_{\mu
\nu }{}^\lambda +T_{\nu \mu }{}^\lambda \right) ,  \nonumber \\
T^\lambda {}_{\mu \nu } &=&e_I{}^\rho T^I{}_{\mu \nu }=\Gamma ^\lambda
{}_{\mu \nu }-\Gamma ^\lambda {}_{\nu \mu }.
\end{eqnarray}
Accordingly the curvature can be represented as
\begin{eqnarray}
R^\rho {}_{\sigma \mu \nu } &=&e_I{}^\rho e^J{}_\sigma R^I{}_{J\mu \nu
}=\partial _\mu \Gamma ^\rho {}_{\sigma \nu }-\partial _\nu \Gamma ^\rho
{}_{\sigma \mu }+\Gamma ^\rho {}_{\lambda \mu }\Gamma ^\lambda {}_{\sigma
\nu }-\Gamma ^\rho {}_{\lambda \nu }\Gamma ^\lambda {}_{\sigma \mu },
\nonumber \\
&=&R_{\left\{ {}\right\} }^\rho {}_{\sigma \mu \nu }+\partial _\mu K^\rho
{}_{\sigma \nu }-\partial _\nu K^\rho {}_{\sigma \mu }+K^\rho {}_{\lambda
\mu }K^\lambda {}_{\sigma \nu }-K^\rho {}_{\lambda \nu }K^\lambda {}_{\sigma
\mu }  \nonumber \\
&&+\left\{ _\lambda {}^\rho {}_\mu \right\} K^\lambda {}_{\sigma \nu
}-\left\{ _\lambda {}^\rho {}_\nu \right\} K^\lambda {}_{\sigma \mu
}+\left\{ _\sigma {}^\lambda {}_\nu \right\} K^\rho {}_{\lambda \mu
}-\left\{ _\sigma {}^\lambda {}_\mu \right\} K^\rho {}_{\lambda \nu },
\end{eqnarray}
where
\[
R_{\left\{ {}\right\} }^\rho {}_{\sigma \mu \nu }=\partial _\mu \left\{
_\sigma {}^\rho {}_\nu \right\} -\partial _\nu \left\{ _\sigma {}^\rho
{}_\mu \right\} +\left\{ _\lambda {}^\rho {}_\mu \right\} \left\{ _\sigma
{}^\lambda {}_\nu \right\} -\left\{ _\lambda {}^\rho {}_\nu \right\} \left\{
_\sigma {}^\lambda {}_\mu \right\} ,
\]
is the curvature of the Christoffel symbol.

\section{Cosmological equations}

For the space flat Friedmann-Robertson-Walker metric
\begin{equation}
g_{\mu \nu }=\text{diag}\left( -1,a\left( t\right) ^2,a\left( t\right)
^2,a\left( t\right) ^2\right) ,
\end{equation}
we have
\begin{eqnarray}
\left\{ _0{}^0{}_0\right\} &=&0,\left\{ _0{}^0{}_i\right\} =\left\{
_i{}^0{}_0\right\} =0,\left\{ _i{}^0{}_j\right\} =a\stackrel{\cdot }{a}%
\delta _{ij},  \nonumber \\
\left\{ _0{}^i{}_0\right\} &=&0,\left\{ _j{}^i{}_0\right\} =\left\{
_0{}^i{}_j\right\} =\frac{\stackrel{\cdot }{a}}a\delta _j^i,\left\{
_j{}^i{}_k\right\} =0,i,j,k,...=1,2,3.
\end{eqnarray}
The non-vanishing torsion components with holonomic indices are given by two
functions $h$ and $f$ [13]:
\begin{eqnarray}
T_{110} &=&T_{220}=T_{330}=a^2h,  \nonumber \\
T_{123} &=&T_{231}=T_{312}=a^3f,
\end{eqnarray}
and then contortion components are
\begin{eqnarray}
K^1{}_{10} &=&K^2{}_{20}=K^3{}_{30}=0,  \nonumber \\
K^1{}_{01} &=&K^2{}_{02}=K^3{}_{03}=h,  \nonumber \\
K^0{}_{11} &=&K^0{}_{22}=K^0{}_{22}={}a^2h,  \nonumber \\
K^1{}_{23} &=&K^2{}_{31}=K^3{}_{12}=-\frac 12af,  \nonumber \\
K^1{}_{32} &=&K^2{}_{13}=K^3{}_{21}=\frac 12af.
\end{eqnarray}
Among the torsion components, only the pseudotrace axial ingredient given by
$f$ couples to spinors in a minimal way. The scalar mode $h$ of\ torsion
could be considered as a ''phantom'' field, at least in the matter-dominated
epoch, since it will not interact directly with matter; it only interacts
indirectly via gravitation.

The non-vanishing components of the curvature $R_{\left\{ {}\right\} }^\rho
{}_{\sigma \mu \nu }$ are
\begin{eqnarray}
R_{\left\{ {}\right\} }^0{}_{101} &=&R_{\left\{ {}\right\}
}^0{}_{202}=R_{\left\{ {}\right\} }^0{}_{303}=a^2\left( \stackrel{\cdot }{H}%
+H^2+Hh+\stackrel{\cdot }{h}\right) ,  \nonumber \\
R_{\left\{ {}\right\} }^0{}_{123} &=&-R_{\left\{ {}\right\}
}^0{}_{213}=R_{\left\{ {}\right\} }^0{}_{312}=a^3f\left( H+h\right) ,
\nonumber \\
R_{\left\{ {}\right\} }^1{}_{203} &=&-R_{\left\{ {}\right\}
}^1{}_{302}=R_{\left\{ {}\right\} }^2{}_{301}=-\frac 12a\left( Hf+\stackrel{%
\cdot }{f}\right) ,  \nonumber \\
R_{\left\{ {}\right\} }^1{}_{212} &=&R_{\left\{ {}\right\}
}^1{}_{313}=R_{\left\{ {}\right\} }^2{}_{323}=a^2\left( H^2+2Hh+h^2-\frac 14%
f^2\right) ,
\end{eqnarray}
\begin{eqnarray}
R_{\left\{ {}\right\} }{}_{00} &=&-3\stackrel{\cdot }{H}-3\stackrel{\cdot }{h%
}-3H^2-3Hh,  \nonumber \\
R_{\left\{ {}\right\} }{}_{11} &=&a^2\left( \stackrel{\cdot }{H}+3H^2+5Hh+%
\stackrel{\cdot }{h}+2h^2-\frac 12f^2\right) ,
\end{eqnarray}
\begin{equation}
R_{\left\{ {}\right\} }{}=6\stackrel{\cdot }{H}+12H^2+18Hh+6\stackrel{\cdot
}{h}+6h^2-\frac 32f^2,
\end{equation}
where $H=\stackrel{\cdot }{a}\left( t\right) /a\left( t\right) $ is the
Hubble parameter. Using these results and (16---20) we can compute

\begin{eqnarray}
e_{I0}\frac{\delta {\cal L}}{\delta e_I{}^0} &=&l^{-2}\{\left( \beta
+3\alpha \right) [-12\left( \stackrel{\cdot }{H}+\stackrel{\cdot }{h}\right)
^2-24\left( \stackrel{\cdot }{H}+\stackrel{\cdot }{h}\right) H\left(
H+h\right)  \nonumber \\
&&+12h\left( h+2H\right) \left( h+H\right) ^2-6\left( h+H\right)
^2f^2+\allowbreak \frac 34f^4]  \nonumber \\
&&+\gamma \left( 18h^2+6f^2\right) +3H^2+6Hh+3h^2-\frac 34f^2-12l^{-2}\}e,
\end{eqnarray}
\begin{eqnarray}
e_{I1}\frac{\delta {\cal L}}{\delta e_I{}^1} &=&-l^{-2}\allowbreak
a^2\{\left( \beta +3\alpha \right) [-4\left( \stackrel{\cdot }{H}+\stackrel{%
\cdot }{h}\right) ^2-8\left( \stackrel{\cdot }{H}+\stackrel{\cdot }{h}%
\right) \left( H^2+Hh\right)  \nonumber \\
&&+\allowbreak 4h\left( h+2H\right) \left( h+H\right) ^2-2\left( h+H\right)
^2f^2+\frac 14f^4]  \nonumber \\
&&-2\gamma \left( 2\stackrel{\cdot }{h}+8Hh+h^2{}+f^2\right) +2\left(
\stackrel{\cdot }{H}+\stackrel{\cdot }{h}\right) +3H^2  \nonumber \\
&&+4Hh+h^2-\frac 14f^2-12l^{-2}\}e,
\end{eqnarray}
\begin{eqnarray}
\frac{\delta {\cal L}}{\delta \Gamma ^{\ 01}{}_1} &=&-2a^{-1}l^{-2}\{\left(
\beta +6\alpha \right) \left( \stackrel{\cdot \cdot }{H}+\stackrel{\cdot
\cdot }{h}\right) +3\left( \beta +4\alpha \right) \left( hH^2+2H\stackrel{%
\cdot }{H}+2h\stackrel{\cdot }{H}\right)  \nonumber \\
&&+\left( \allowbreak 5\beta +18\alpha \right) \left( H\stackrel{\cdot }{h}+h%
\stackrel{\cdot }{h}+h^2H\right) +\left( \beta +3\alpha \right) \left( 2h^3-f%
\stackrel{\cdot }{f}-\frac 12hf^2\right) +\frac 14h\}e,
\end{eqnarray}
\begin{eqnarray}
&&\frac{\delta {\cal L}}{\delta \Gamma ^{12}{}_3}=a^{-1}l^{-2}f\{2\left(
\beta +6\alpha \right) \left( \stackrel{\cdot }{H}+\stackrel{\cdot }{h}%
\right) +6\left( \beta +\allowbreak 4\alpha \right) H^2  \nonumber \\
&&+2\left( 5\beta +18\alpha \right) Hh+\left( \beta \ +3\alpha \right)
\left( 4h^2-f^2\right)  \nonumber \\
&&-4\gamma +\frac 12\}e,
\end{eqnarray}

Suppose the matter source is a fluid characterized by the density $\rho $
the pressure $p$ and the spin $s_{IJ}{}^\mu $. The system of field equations
(15) consists of four independent ones:
\begin{eqnarray}
e_{I0}\frac{\delta {\cal L}}{\delta e_I{}^0} &=&-e_{I0}\frac{\delta {\cal L}%
_\psi }{\delta e_I{}^0}=\rho ,  \nonumber \\
e_{I1}\frac{\delta {\cal L}}{\delta e_I{}^1} &=&-e_{I1}\frac{\delta {\cal L}%
_\psi }{\delta e_I{}^1}=g_{11}p,  \nonumber \\
\frac{\delta {\cal L}}{\delta \Gamma {}^{01}{}_1} &=&-\frac{\delta {\cal L}%
_\psi }{\delta \Gamma {}^{01}{}_1}=e_1{}^1s_{01}{}^1,  \nonumber \\
\frac{\delta {\cal L}}{\delta \Gamma {}^{12}{}_3} &=&-\frac{\delta {\cal L}%
_\psi }{\delta \Gamma {}^{12}{}_3}=e_3{}^3s_{12}{}^3.
\end{eqnarray}
Using (28-31) the Lagrange equations (32) can be written as
\begin{eqnarray}
&&\left( \beta +3\alpha \right) [-12\left( \stackrel{\cdot }{H}+\stackrel{%
\cdot }{h}\right) ^2-24\left( \stackrel{\cdot }{H}+\stackrel{\cdot }{h}%
\right) H\left( H+h\right)  \nonumber \\
&&+12h\left( h+2H\right) \left( h+H\right) ^2-6\left( h+H\right)
^2f^2+\allowbreak \frac 34f^4]  \nonumber \\
&&+\gamma \left( 18h^2+6f^2\right) +3H^2+6Hh+3h^2-\frac 34%
f^2-12l^{-2}-l^2\rho =0,
\end{eqnarray}
\begin{eqnarray}
&&\left( \beta +3\alpha \right) [-4\left( \stackrel{\cdot }{H}+\stackrel{%
\cdot }{h}\right) ^2-8\left( \stackrel{\cdot }{H}+\stackrel{\cdot }{h}%
\right) \left( H^2+Hh\right)  \nonumber \\
&&+\allowbreak 4h\left( h+2H\right) \left( h+H\right) ^2-2\left( h+H\right)
^2f^2+\frac 14f^4]  \nonumber \\
&&+2\gamma \left( 2\stackrel{\cdot }{h}+8Hh+h^2{}+f^2\right) -2\left(
\stackrel{\cdot }{H}+\stackrel{\cdot }{h}\right) -3H^2  \nonumber \\
&&-4Hh-h^2+\frac 14f^2+12l^{-2}+l^2p=0,
\end{eqnarray}
\begin{eqnarray}
&&-2\{\left( \beta +6\alpha \right) \left( \stackrel{\cdot \cdot }{H}+%
\stackrel{\cdot \cdot }{h}\right) +3\left( \beta +4\alpha \right) \left(
hH^2+2H\stackrel{\cdot }{H}+2h\stackrel{\cdot }{H}\right) +\left(
\allowbreak 5\beta +18\alpha \right) \left( H\stackrel{\cdot }{h}+h\stackrel{%
\cdot }{h}+h^2H\right)  \nonumber \\
&&+\left( \beta +3\alpha \right) \left( 2h^3-f\stackrel{\cdot }{f}-\frac 12%
hf^2\right) +\frac 14h\}-l^2s_{01}{}^1=0,
\end{eqnarray}
\begin{eqnarray}
&&f\{2\left( \beta +6\alpha \right) \left( \stackrel{\cdot }{H}+\stackrel{%
\cdot }{h}\right) +6\left( \beta +\allowbreak 4\alpha \right) H^2  \nonumber
\\
&&+2\left( 5\beta +18\alpha \right) Hh+\left( \beta \ +3\alpha \right)
\left( 4h^2-f^2\right) -4\gamma +\frac 12\}-l^2s_{12}{}^3=0.
\end{eqnarray}
Assuming $s_{\mu \nu }{}^\lambda $ $=0$ (i.e., the source spin current is
negligible), the Eq. (36) reads

\begin{eqnarray}
&&f\{2\left( \beta +6\alpha \right) \left( \stackrel{\cdot }{H}+\stackrel{%
\cdot }{h}\right) +6\left( \beta +\allowbreak 4\alpha \right) H^2  \nonumber
\\
&&+2\left( 5\beta +18\alpha \right) Hh+\left( \beta \ +3\alpha \right)
\left( 4h^2-f^2\right) -4\gamma +\frac 12\}=0,
\end{eqnarray}
and gives

\begin{equation}
f=0,
\end{equation}
or
\begin{eqnarray}
&&2\left( \beta +6\alpha \right) \left( \stackrel{\cdot }{H}+\stackrel{\cdot
}{h}\right) +6\left( \beta +\allowbreak 4\alpha \right) H^2  \nonumber \\
&&+2\left( 5\beta +18\alpha \right) Hh+\left( \beta \ +3\alpha \right)
\left( 4h^2-f^2\right) -4\gamma +\frac 12=0.
\end{eqnarray}
Therefore, we have two cases.

In the first case, $f=0$, the Eqs (33) and (34) read
\begin{eqnarray}
&&\left( \beta +3\alpha \right) [-\left( \stackrel{\cdot }{H}+\stackrel{%
\cdot }{h}\right) ^2-2\left( \stackrel{\cdot }{H}+\stackrel{\cdot }{h}%
\right) H\left( H+h\right) +h\left( h+2H\right) \left( h+H\right) ^2]
\nonumber \\
&&+\frac 32\gamma h^2+\frac 14H^2+\frac 12Hh+\frac 14h^2-l^{-2}-\frac{%
l^2\rho }{12}=0,
\end{eqnarray}
and
\begin{eqnarray}
&&\left( \beta +3\alpha \right) [-\left( \stackrel{\cdot }{H}+\stackrel{%
\cdot }{h}\right) ^2-2\left( \stackrel{\cdot }{H}+\stackrel{\cdot }{h}%
\right) \left( H^2+Hh\right) +h\left( h+2H\right) \left( h+H\right) ^2]
\nonumber \\
&&+\frac 12\gamma \left( 2\stackrel{\cdot }{h}+8Hh+h^2{}\right) -\frac 12%
\left( \stackrel{\cdot }{H}+\stackrel{\cdot }{h}\right) -\frac 34H^2-Hh-%
\frac{h^2}4+3l^{-2}+\frac{l^2p}4=0,
\end{eqnarray}
which lead to
\begin{equation}
\stackrel{\cdot }{H}=\left( 2\gamma -1\right) \stackrel{\cdot }{h}%
-2H^2+\left( 8\gamma -3\right) Hh{}-\left( 2\gamma +1\right) h^2+8l^{-2}+%
\frac{l^2}6\left( \rho +3p\right) ,
\end{equation}
and
\begin{eqnarray}
&&-4\gamma ^2\stackrel{\cdot }{h}^2+\gamma \left( 4H^2+8\left( 1-4\gamma
\right) Hh+4\left( 2\gamma +1\right) h^2-\frac 23l^2\left( \rho -3p\right) -%
\frac{32}{l^2}\right) \stackrel{\cdot }{h}  \nonumber \\
&&+16H^3h\gamma +\left( 28\gamma -64\gamma ^2\right) H^2h^2+8\left( 4\gamma
+1\right) \gamma h^3H-4\left( \gamma +1\right) \gamma h^4  \nonumber \\
&&+\left( \frac{16}{l^2}+\frac 13l^2\left( \rho -3p\right) \right) H^2+\frac %
1{4\left( \beta +3\alpha \right) }H^2+\left( 1-4\gamma \right) \left( \frac{%
32}{l^2}+\allowbreak \frac 23l^2\left( \rho -3p\right) \right) Hh  \nonumber
\\
&&+\frac 1{2\left( \beta +3\alpha \right) }Hh+\left( 1+2\gamma \right)
\left( \frac{16}{l^2}+\allowbreak \frac 13l^2\left( \rho -3p\right) \right)
h^2+\frac{\left( 6\gamma +1\right) }{4\left( \beta +3\alpha \right) }h^2
\nonumber \\
&&-\frac{l^2\rho }{12\left( \beta +3\alpha \right) }-\frac 83\left( \rho
-3p\right) -\allowbreak \frac 1{36}l^4\left( \rho -3p\right) ^2-\frac 1{%
\left( \beta +3\alpha \right) l^2}-\frac{64}{l^4}  \nonumber \\
&=&0.
\end{eqnarray}

So we have the equations (42), (43) and
\begin{eqnarray}
&&\left( \beta +6\alpha \right) \left( \stackrel{\cdot \cdot }{H}+\stackrel{%
\cdot \cdot }{h}\right) +3\left( \beta +4\alpha \right) \left( hH^2+2H%
\stackrel{\cdot }{H}+2h\stackrel{\cdot }{H}\right) +\left( \allowbreak
5\beta +18\alpha \right) \left( H\stackrel{\cdot }{h}+h\stackrel{\cdot }{h}%
+h^2H\right)  \nonumber \\
&&+2\left( \beta +3\alpha \right) h^3+\frac 14h=0,
\end{eqnarray}
for the unknown functions $H$ and $h$. $\allowbreak $

In the second case, $f$ satisfies the condition (39). The Eqs. (33) and (34)
yield
\begin{equation}
\stackrel{\cdot }{H}=\left( 2\gamma -1\right) \stackrel{\cdot }{h}%
{}-2H^2+\left( 8\gamma -3\right) Hh-\left( 2\gamma +1\right) h^2+\frac 14%
f^2+8l^{-2}+\frac{l^2}6\left( \rho +3p\right) ,
\end{equation}
and
\begin{eqnarray}
&&-4\gamma ^2\stackrel{\cdot }{h}^2+\gamma \left( 4H^2+8\left( 1-4\gamma
\right) Hh+4\left( 1+2\gamma \right) h^2-f^2-\frac 23l^2B-\frac{32}{l^2}%
\right) \stackrel{\cdot }{h}  \nonumber \\
&&+16H^3h\gamma +4\gamma \left( 7-16\gamma \right) H^2h^2+8\gamma \left(
1+4\gamma \right) h^3H-4\gamma \left( 1+\gamma \right) h^4-\allowbreak
4\gamma Hhf^2+\gamma h^2f^2  \nonumber \\
&&+\left( \frac{16}{l^2}+\frac 13l^2\left( \rho +3p\right) \right) H^2+\frac %
1{4\left( \beta +3\alpha \right) }H^2+\left( 1-\allowbreak 4\gamma \right)
\left( \frac{32}{l^2}+\allowbreak \frac 23l^2B\right) Hh+\frac 1{2\left(
\beta +3\alpha \right) }Hh  \nonumber \\
&&+\frac{16}{l^2}\left( 1+2\gamma \right) h^2+\allowbreak \frac 13\left(
1+2\gamma \right) l^2\left( \rho +3p\right) h^2+\frac{6\gamma +1}{4\left(
\beta +3\alpha \right) }h^2-\left( \frac 4{l^2}+\frac 1{12}l^2\left( \rho
+3p\right) \right) f^2  \nonumber \\
&&+\frac{8\gamma -1}{16\left( \beta +3\alpha \right) }f^2-\frac 1{\left(
\beta +3\alpha \right) }l^{-2}-\frac{l^2}{12\left( \beta +3\alpha \right) }%
\rho -\frac 83\left( \rho +3p\right) -\allowbreak \frac 1{36}l^4\left( \rho
+3p\right) ^2-\frac{64}{l^4}  \nonumber \\
&=&0.
\end{eqnarray}
$\allowbreak $

The Eqs. (45) and (39) gives
\begin{eqnarray}
f^2 &=&8\gamma \frac{\left( \beta +6\alpha \right) }\beta \stackrel{\cdot }{h%
}{}+4H^2+8\left( 1+4\gamma \frac{\left( \beta +6\alpha \right) }\beta
\right) Hh+\left( 4-8\gamma \frac{\left( \beta +6\alpha \right) }\beta
\right) h^2  \nonumber \\
&&+\frac{1-8\gamma }\beta +\frac{32\left( \beta +6\alpha \right) }{\beta l^2}%
+\frac{2\left( \beta +6\alpha \right) }{3\beta }l^2\left( \rho +3p\right) .
\end{eqnarray}

Substituting into (45) and (46) yields
\begin{eqnarray}
&&-12\gamma ^2\frac{\beta +4\alpha }\beta \stackrel{\cdot }{h}^2  \nonumber
\\
&&+\left( -32\frac \gamma \beta \left( \gamma \allowbreak \left( \beta
+6\alpha \right) -2\left( \beta +3\alpha \right) \right) Hh+8\frac \gamma %
\beta \left( \gamma \left( \beta +6\alpha \right) +2\left( \beta +3\alpha
\right) \right) h^2\right) \stackrel{\cdot }{h}  \nonumber \\
&&+\left( \frac{8\gamma -1}\beta \left( \frac{\gamma \left( \beta +6\alpha
\right) }{2\left( \beta +3\alpha \right) }+1\right) -\frac 4{l^2}\allowbreak
\frac{17\beta +48\alpha }\beta -\frac{17\beta +48\alpha }{12\beta }l^2\left(
\rho +3p\right) \right) \stackrel{\cdot }{h}  \nonumber \\
&&-192\gamma ^2\frac{\beta +4\alpha }\beta H^2h^2+96\gamma ^2\frac{\beta
+4\alpha }\beta Hh^3-12\gamma ^2\frac{\beta +4\alpha }\beta h^4{}\allowbreak
\nonumber \\
&&+\frac{2\gamma }{\beta +3\alpha }H^2+\left[ 2\gamma \frac{24\gamma \beta
+96\gamma \alpha -\beta -12\alpha }{\beta \left( \beta +3\alpha \right) }%
-384\gamma \frac{\beta +4\alpha }{\beta l^2}\allowbreak -8\gamma \frac{\beta
+4\alpha }\beta l^2\left( \rho +3p\right) \right] Hh  \nonumber \\
&&+\left[ -\gamma \frac{-5\beta +12\gamma \beta +48\gamma \alpha -6\alpha }{%
\beta \left( \beta +3\alpha \right) }+96\gamma \frac{\beta +4\alpha }{\beta
l^2}+2\gamma \frac{\beta +4\alpha }\beta l^2\left( \rho +3p\right) \right]
h^2  \nonumber \\
&&+\frac{1-8\gamma }\beta \frac{8\gamma -1}{16\left( \beta +3\alpha \right) }%
+\frac{48\gamma \beta +192\gamma \alpha -7\beta -24\alpha }{\beta \left(
\beta +3\alpha \right) l^2}-192\frac{\beta +4\alpha }{\beta l^4}  \nonumber
\\
&&-\frac{l^2}{12\left( \beta +3\alpha \right) }\rho +\left( \frac{8\gamma -1}%
8\frac{\beta +4\alpha }{\beta \left( \beta +3\alpha \right) }l^2-8\frac{%
\beta +4\alpha }\beta \right) \left( \rho +3p\right) -\frac{\beta +4\alpha }{%
12\beta }l^4\left( \rho +3p\right) ^2  \nonumber \\
&=&0.
\end{eqnarray}
and
\begin{eqnarray}
\stackrel{\cdot }{H} &=&\left( \frac{4\gamma \left( \beta +3\alpha \right) }%
\beta -1\right) \stackrel{\cdot }{h}{}-H^2+\left( 16\gamma \frac{\beta
+3\alpha }\beta -1\right) Hh-4\gamma \frac{\beta +3\alpha }\beta h^2
\nonumber \\
&&+\frac{1-8\gamma }{4\beta }+\allowbreak 16\frac{\beta +3\alpha }{\beta l^2}%
+\frac{\beta +3\alpha }{3\beta }l^2\left( \rho -3p\right) .
\end{eqnarray}
Differentiating (47) gives
\begin{eqnarray*}
-f\stackrel{\cdot }{f} &=&-4\gamma \frac{\left( \beta +6\alpha \right) }\beta
\stackrel{\cdot \cdot }{h}{}-4\left( 1+4\gamma \frac{\left( \beta +6\alpha
\right) }\beta \right) h\stackrel{\cdot }{H}-4H\stackrel{\cdot }{H}%
{}-4\left( 1+4\gamma \frac{\left( \beta +6\alpha \right) }\beta \right) H%
\stackrel{\cdot }{h}{}{}-\left( 4-8\gamma \frac{\left( \beta +6\alpha
\right) }\beta \right) h\stackrel{\cdot }{h}{} \\
&&-\frac{\left( \beta +6\alpha \right) }{3\beta }l^2\left( \stackrel{\cdot }{%
\rho }+3\stackrel{\cdot }{p}\right) .
\end{eqnarray*}
Substituting it and (47) into (35) and letting $s_{01}{}^1=0$ give
\begin{eqnarray}
&&\stackrel{\cdot \cdot }{H}+\left( 1-4\gamma \frac{\left( \beta +3\alpha
\right) }\beta \right) \stackrel{\cdot \cdot }{h}+\allowbreak 2H\stackrel{%
\cdot }{H}+\allowbreak 2\left( 1-\frac{8\gamma }\beta \left( \beta +3\alpha
\right) \right) h\stackrel{\cdot }{H}  \nonumber \\
&&+\left( 1-16\gamma \frac{\left( \beta +3\alpha \right) }\beta \right) H%
\stackrel{\cdot }{h}{}{}+\left( 4\gamma \frac{\left( \beta +3\alpha \right) }%
\beta +1\right) h\stackrel{\cdot }{h}{}  \nonumber \\
&&+hH^2+\left( 1-16\gamma \frac{\left( \beta +3\alpha \right) }\beta \right)
Hh^2+4\gamma \frac{\left( \beta +3\alpha \right) }\beta h^3  \nonumber \\
&&+\left( \allowbreak \allowbreak 4\gamma \frac{\beta +3\alpha }{\beta
\left( \beta +6\alpha \right) }\allowbreak -\frac 1{4\beta }\right) h-\frac{%
16\left( \beta +3\alpha \right) }{\beta l^2}h-\frac{\left( \beta +3\alpha
\right) }{3\beta }l^2h\left( \rho +3p\right)  \nonumber \\
&&-\frac{\left( \beta +3\alpha \right) }{3\beta }l^2\left( \stackrel{\cdot }{%
\rho }+3\stackrel{\cdot }{p}\right)  \nonumber \\
&=&0.
\end{eqnarray}

So we have the equations (48), (49), and (50) for the unknown functions $H$
and $h$. The unknown function $f$ is given by (47).

\section{Two specific models}

In order to emphasize the geometrical nature of the effect of acceleration
of cosmological expansion we concentrate on vacuum solutions in two specific
cases and discuss only the acceleration solutions.

\subsection{When $\beta =-3\alpha $}

This corresponds to conformal (Weyl) gravity which has been investigated by
numerous authors [recent, see 2 and 9] but it must be pointed out that the
principle and structure between the theory here and higher-derivative
gravity in Mannheim's theory are quite\ different.

According to last section, the equation (37) gives two cases.

In the first case $f=0$, the functions $H$ and $h$ now satisfy the equations
(40), (41) and (44), i.e.,
\begin{equation}
\left( 6\gamma +1\right) h^2+H^2+2Hh-4l^{-2}=0,
\end{equation}
\begin{equation}
\left( 4\gamma -2\right) \stackrel{\cdot }{h}-2\stackrel{\cdot }{H}+\left(
16\gamma -4\right) Hh+\left( 2\gamma {}-1\right) h^2-3H^2+12l^{-2}=0,
\end{equation}
\begin{eqnarray}
&&\stackrel{\cdot \cdot }{H}+\stackrel{\cdot \cdot }{h}+2\left( H+h\right)
\stackrel{\cdot }{H}  \nonumber \\
&&+\left( H+h\right) \stackrel{\cdot }{h}+hH^2+h^2H+\frac 1{12\alpha }h=0.
\end{eqnarray}

Eq. (51) has the roots
\begin{equation}
h=\frac{-H\pm \sqrt{-6\gamma H^2+4\left( 6\gamma +1\right) l^{-2}}}{\left(
6\gamma +1\right) }.
\end{equation}
Eq. (52) gives
\begin{equation}
\stackrel{\cdot }{h}=\frac 1{\left( 2\gamma -1\right) }\stackrel{\cdot }{H}-%
\frac{\left( 8\gamma -2\right) }{\left( 2\gamma -1\right) }Hh-\frac 12h^2+%
\frac 3{2\left( 2\gamma -1\right) }H^2-\frac{6l^{-2}}{\left( 2\gamma
-1\right) },
\end{equation}
and then
\begin{equation}
\stackrel{\cdot \cdot }{h}=\frac 1{\left( 2\gamma -1\right) }\stackrel{\cdot
\cdot }{H}-\frac{\left( 8\gamma -2\right) }{\left( 2\gamma -1\right) }h%
\stackrel{\cdot }{H}-\left( \frac{\left( 8\gamma -2\right) }{\left( 2\gamma
-1\right) }H+\frac 12h\right) \stackrel{\cdot }{h}+\frac 3{2\left( 2\gamma
-1\right) }H\stackrel{\cdot }{H}.
\end{equation}
Substituting (54), (55) and (56) into (53) yields
\begin{eqnarray}
\stackrel{\cdot \cdot }{H} &=&-\left( \frac{48\gamma ^3-50\gamma ^2-7\gamma
+2}{2\gamma \left( 2\gamma -1\right) \left( 6\gamma +1\right) }H\mp \frac{%
8\gamma -1}{4\gamma }\frac{\sqrt{-6\gamma H^2+4\left( 6\gamma +1\right)
l^{-2}}}{\left( 6\gamma +1\right) }\right) \stackrel{\cdot }{H}+\frac{%
2\left( 504\gamma ^3+324\gamma ^2-26\gamma -3\right) }{\left( 6\gamma
+1\right) ^3\left( 2\gamma -1\right) }H^3  \nonumber \\
&&+\allowbreak \frac{840\gamma ^3-4\gamma ^2-6\gamma -5}{\left( 2\gamma
-1\right) \left( 6\gamma +1\right) ^2\gamma l^2}H+\frac{2\gamma -1}{24\alpha
\gamma \left( 6\gamma +1\right) }H  \nonumber \\
&&\mp \left( \frac{-476\gamma ^2+2592\gamma ^4+744\gamma ^3-18\gamma +1}{%
4\gamma \left( 6\gamma +1\right) ^3\left( 2\gamma -1\right) }H^2+\frac{%
3\left( 2\gamma +1\right) }{\gamma \left( 6\gamma +1\right) ^2l^2}+\frac{%
2\gamma -1}{24\alpha \gamma \left( 6\gamma +1\right) }\right) \sqrt{-6\gamma
H^2+4\left( 6\gamma +1\right) l^{-2}},
\end{eqnarray}

Let
\[
\stackrel{\cdot }{H}=X.
\]
We have the dynamical system

\begin{eqnarray}
\stackrel{\cdot }{H} &=&X,  \nonumber \\
\stackrel{\cdot }{X} &=&-\left( \frac{48\gamma ^3-50\gamma ^2-7\gamma +2}{%
2\gamma \left( 2\gamma -1\right) \left( 6\gamma +1\right) }H\mp \frac{%
8\gamma -1}{4\gamma }\frac{\sqrt{-6\gamma H^2+4\left( 6\gamma +1\right)
l^{-2}}}{\left( 6\gamma +1\right) }\right) X  \nonumber \\
&&+AH^3+\allowbreak BH\mp (CH^2+D)\sqrt{-6\gamma H^2+4\left( 6\gamma
+1\right) l^{-2}}
\end{eqnarray}
where
\begin{eqnarray}
A &=&\frac{2\left( 504\gamma ^3+324\gamma ^2-26\gamma -3\right) }{\left(
6\gamma +1\right) ^3\left( 2\gamma -1\right) },  \nonumber \\
B &=&\frac{840\gamma ^3-4\gamma ^2-6\gamma -5}{\left( 2\gamma -1\right)
\left( 6\gamma +1\right) ^2\gamma l^2}+\frac{2\gamma -1}{24\alpha \gamma
\left( 6\gamma +1\right) },  \nonumber \\
C &=&\frac{-476\gamma ^2+2592\gamma ^4+744\gamma ^3-18\gamma +1}{4\gamma
\left( 6\gamma +1\right) ^3\left( 2\gamma -1\right) },  \nonumber \\
D &=&\frac{3\left( 2\gamma +1\right) }{\gamma \left( 6\gamma +1\right) ^2l^2}%
+\frac{2\gamma -1}{24\alpha \gamma \left( 6\gamma +1\right) }.
\end{eqnarray}
The Jacobian elements are
\[
\frac{\partial \stackrel{\cdot }{H}}{\partial H}=0,\frac{\partial \stackrel{%
\cdot }{H}}{\partial X}=1,
\]
\begin{eqnarray*}
\frac{\partial \stackrel{\cdot }{X}}{\partial H} &=&\left( -\frac{48\gamma
^3-50\gamma ^2-7\gamma +2}{2\gamma \left( 2\gamma -1\right) \left( 6\gamma
+1\right) }\mp \frac{3\left( 8\gamma -1\right) H}{2\left( 6\gamma +1\right)
\sqrt{-6\gamma H^2+4\left( 6\gamma +1\right) l^{-2}}}\right) X \\
&&+AH^3+\allowbreak BH\mp (CH^2+D)\sqrt{-6\gamma H^2+4\left( 6\gamma
+1\right) l^{-2}},
\end{eqnarray*}
\begin{equation}
\frac{\partial \stackrel{\cdot }{X}}{\partial X}=-\frac{48\gamma ^3-50\gamma
^2-7\gamma +2}{2\gamma \left( 2\gamma -1\right) \left( 6\gamma +1\right) }%
H\pm \frac{8\gamma -1}{4\gamma \left( 6\gamma +1\right) }\sqrt{-6\gamma
H^2+4\left( 6\gamma +1\right) l^{-2}}.
\end{equation}

The critical point equations are

\begin{eqnarray}
X &=&0,  \nonumber \\
AH^3+\allowbreak BH\mp \left( CH^2+D\right) \sqrt{-6\gamma H^2+4\left(
6\gamma +1\right) l^{-2}} &=&0.
\end{eqnarray}
Rationalization gives

\begin{equation}
H^6+aH^4+b\allowbreak H^2\allowbreak +c=0,
\end{equation}
where

\begin{eqnarray}
a &=&\frac{\allowbreak 2\left( l^2AB-12C^2\gamma -2C^2+6CD\gamma l^2\right)
}{\allowbreak \allowbreak \left( A^2+6C^2\gamma \right) l^2},  \nonumber \\
b &=&\frac{B^2l^2-48\gamma DC-8CD+6D^2\gamma l^2}{\allowbreak \allowbreak
\left( A^2+6C^2\gamma \right) l^2},  \nonumber \\
c &=&-\frac{4\left( 6\gamma +1\right) }{\allowbreak \allowbreak \left(
A^2+6C^2\gamma \right) l^2}D^2.
\end{eqnarray}
The equation (62) has the roots

\begin{eqnarray}
H_1^2 &=&\left( -\frac q2+\sqrt{\Delta }\right) ^{1/3}+\left( -\frac q2-%
\sqrt{\Delta }\right) ^{1/3}-\frac a3,  \nonumber \\
H_2^2 &=&\left( -\frac q2+\sqrt{\Delta }\right) ^{1/3}\omega +\left( -\frac q%
2-\sqrt{\Delta }\right) ^{1/3}\omega ^2-\frac a3,  \nonumber \\
H_3^2 &=&\left( -\frac q2+\sqrt{\Delta }\right) ^{1/3}\omega ^2+\left( -%
\frac q2-\sqrt{\Delta }\right) ^{1/3}\omega -\frac a3.
\end{eqnarray}
where
\begin{eqnarray*}
p &=&\left( -\frac 13a^2+b\right) , \\
q &=&\frac 2{27}a^3-\frac 13ba+c,
\end{eqnarray*}
\begin{equation}
\Delta =\left( \frac q2\right) ^2+\left( \frac p3\right) ^3,
\end{equation}
and
\begin{equation}
\omega =\frac{-1+\sqrt{3}i}2.
\end{equation}

Now we have the critical points
\begin{eqnarray*}
H_1 &=&\pm \sqrt{\left( -\frac q2+\sqrt{\Delta }\right) ^{1/3}+\left( -\frac %
q2-\sqrt{\Delta }\right) ^{1/3}-\frac a3},X_1=0, \\
H_2 &=&\pm \sqrt{\left( -\frac q2+\sqrt{\Delta }\right) ^{1/3}\omega +\left(
-\frac q2-\sqrt{\Delta }\right) ^{1/3}\omega ^2-\frac a3},X_2=0, \\
H_3 &=&\pm \sqrt{\left( -\frac q2+\sqrt{\Delta }\right) ^{1/3}\omega
^2+\left( -\frac q2-\sqrt{\Delta }\right) ^{1/3}\omega -\frac a3},X_3=0.
\end{eqnarray*}

In order to analyze their stability we give the parameter $\alpha $ and $%
\gamma $ specific values and then obtain the results:

When
\[
\alpha =\frac 1{32}l^2,\gamma =-\frac 14,
\]
the equations (62) become
\[
431H^6l^6-13700H^4l^4+15798H^2l^2+3600=0.
\]
It has the roots
\[
H^2\approx 1.\,4024/l^2,H^2\approx -.\,19478/l^2-2.0\times
10^{-9}i/l^2,H^2\approx 10.\,596/l^2+.\,92212i/l^2,
\]
the first root $H^2=1.\,4024/l^2$ corresponds a positive critical point
\[
H=1.\,1842/l,X=0.
\]

At this point, for

\[
h=\frac{-H+\sqrt{-6\gamma H^2+4\left( 6\gamma +1\right) l^{-2}}}{\left(
6\gamma +1\right) }=\frac{1.\,725}l,
\]
the dynamic system (58) reads
\begin{eqnarray*}
\stackrel{\cdot }{H} &=&X, \\
\stackrel{\cdot }{X} &=&-\left( \frac 13H+3\sqrt{\left( 6H^2-\frac 8{l^2}%
\right) }\right) X \\
&&+\frac{508}3H^3-\frac{196}{l^2}H-(\frac{412}3H^2-\frac{40}{l^2})\sqrt{%
\frac 32H^2-\frac 2{l^2}}.
\end{eqnarray*}

The Jacobian
\[
{\cal M}=\left(
\begin{array}{ll}
0 & 1 \\
-\frac{430.\,76}{l^2} & -\frac{2.\,325}l
\end{array}
\right)
\]
has the eigenvalues $-1.1625/l-20.7222i/l$, $-1.1625/l+20.7222i/l$.
Therefore, the critical point
\[
H=1.\,1842/l,X=0,
\]
is stable, where
\[
h=\frac{1.\,725}l,f=0.
\]

When
\[
\alpha =\left( \frac 1{32}l^2\right) ,\gamma =\left( \frac 14\right) ,
\]
the equations (62) become
\[
1827H^6l^6+5226H^4l^4+2579H^2l^2-2312=0.
\]
It has roots
\[
H^2\approx 0.\,4412/l^2,H^2\approx -1.\,6508/l^2+0.\,37826i/l^2,H^2\approx
-1.\,6508/l^2-0.\,37826i/l^2,
\]
the first root $H^2=0.\,4412/l^2$ corresponds a positive critical point
\[
H=0.\,66423/l,X=0.
\]

At this point, for

\[
h=\frac{-H-\sqrt{-6\gamma H^2+4\left( 6\gamma +1\right) l^{-2}}}{\left(
6\gamma +1\right) }=-\frac{1.\,488}l,
\]
the dynamic system (58) reads
\begin{eqnarray*}
\stackrel{\cdot }{H} &=&X, \\
\stackrel{\cdot }{X} &=&-\left( \frac{17}5H+\frac 15\sqrt{\left( -6H^2+\frac{%
40}{l^2}\right) }\right) X \\
&&-\frac{596}{125}H^3-\frac{692}{75l^2}BH+(\frac{184}{125}H^2+\frac{136}{%
75l^2})\sqrt{-\frac 32H^2+\frac{10}{l^2}}.
\end{eqnarray*}
The Jacobian
\[
{\cal M}=\left(
\begin{array}{ll}
0 & 1 \\
-\frac{10.\,365}{l^2} & -\frac{3.\,4807}l
\end{array}
\right)
\]
has the eigenvalues $-1.74035/l-2.70854i/l$, $-1.74035/l+2.70854i/l$.
Therefore, the critical point
\[
H=0.\,66423/l,X=0,
\]
is stable, there
\[
h=-\frac{1.\,488}l,f=0.
\]

For the two points
\[
X=\stackrel{\cdot }{H}=0,
\]
which corresponds to a de Sitter spacetime.

Following Lu and Pope [1], we chose $\alpha =-\frac 1{2\Lambda }$, which
means
\[
\alpha =-\frac{l^2}{48}.
\]
In contrast with them we deal with a de Sitter spacetime with torsion and
the gravitational Lagrangian including a term $\gamma l^{-2}T{}^\mu {}_{\nu
\rho }T{}_\mu {}^{\nu \rho }$. When we chose
\[
\gamma =-\frac 14,
\]
(59) and (63) give, respectively,
\[
A=\frac{508}3,B=-\frac{156}{l^2},C=\frac{412}3,D=0,
\]
and
\[
a=-\frac{17000}{431l^2},b=\frac{27378}{431l^4},c=0.
\]
The dynamical system (58) becomes

\begin{eqnarray*}
\stackrel{\cdot }{H} &=&X, \\
\stackrel{\cdot }{X} &=&-\left( \frac 13H\pm \sqrt{\frac 32H^2-\frac 2{l^2}}%
\right) X+\frac{508}3H^3-\frac{156}{l^2}H\mp \frac{412}3H^2\sqrt{\frac 32H^2-%
\frac 2{l^2}}
\end{eqnarray*}
(62) becomes
\[
\left( H^4-\frac{17000}{431l^2}H^2+\frac{27378}{431l^4}\right) \allowbreak
H^2\allowbreak =0,
\]
and has the roots
\[
H_1=0,H_2=\sqrt{\frac{8500-103\sqrt{5698}}{431}}/l,H_3=\sqrt{\frac{8500+103%
\sqrt{5698}}{431}}/l.
\]
Therefore we get three critical points
\begin{eqnarray*}
H_1 &=&0,X_1=0, \\
H_2 &=&\sqrt{\frac{8500-103\sqrt{5698}}{431}}/l,X_2=0, \\
H_3 &=&\sqrt{\frac{8500+103\sqrt{5698}}{431}}/l,X_3=0.
\end{eqnarray*}
At the point
\[
H_1=0,X_1=0,
\]
the Jacobian
\[
M=\left(
\begin{array}{ll}
0 & 1 \\
-\frac{156}{l^2} & \mp \frac{6\sqrt{2}i}l
\end{array}
\right)
\]
has the eigenvalues
\[
\left( -3\sqrt{2}+\sqrt{174}\right) i/l,\left( -3\sqrt{2}-\sqrt{174}\right)
i/l.
\]
This point is a center.

At the point
\[
H_2=\sqrt{\frac{8500-103\sqrt{5698}}{431}}/l,X_2=0,
\]
the Jacobian
\[
M=\left(
\begin{array}{ll}
0 & 1 \\
-\frac{180.\,44}{l^2} & -\frac{4.\,7728}l
\end{array}
\right)
\]
has the eigenvalues
\[
-2.3864/l+13.2191i/l,-2.3864/l-13.2191i/l.
\]
This is a stable critical point, where
\[
h=1.\,1472/l,f=0.
\]

At the point
\[
H_3=\sqrt{\frac{8500+103\sqrt{5698}}{431}}/l,X_3=0
\]
the Jacobian
\[
M=\left(
\begin{array}{ll}
0 & 1 \\
\frac{84.\,5}{l^2} & -\frac{46.\,4}l
\end{array}
\right)
\]
has the eigenvalues
\[
-48.15/l,1.755/l.
\]
This is a unstable critical point.

If we chose

\[
\gamma =\frac 14,
\]
(59) and (63) give, respectively,
\[
A=-\frac{596}{125},B=-\frac{164}{25l^2},C=\frac{184}{125},D=\frac{112}{25l^2}%
,
\]
and
\[
a=\frac{474}{203l^2},b=-\frac{459}{203l^4},c=-\frac{224}{29l^6}.
\]
The dynamical system (58) becomes

\begin{eqnarray*}
\stackrel{\cdot }{H} &=&X, \\
\stackrel{\cdot }{X} &=&-\left( \frac{17}5H\mp \frac 25\sqrt{-\frac 32H^2+%
\frac{10}{l^2}}\right) X-\frac{596}{125}H^3-\frac{164}{25l^2}H\mp (\frac{184%
}{125}H^2+\frac{112}{25l^2})\sqrt{-\frac 32H^2+\frac{10}{l^2}}.
\end{eqnarray*}
(62) becomes
\[
H^6+\frac{474}{203l^2}H^4-\frac{459}{203l^4}H^2\allowbreak -\frac{224}{29l^6}%
=0
\]
It has a real root
\[
H=1.30134/l.
\]

At the critical point
\[
H=1.30134/l,X=0,
\]
the Jacobian
\[
M=\left(
\begin{array}{ll}
0 & 1 \\
-\frac{36.\,265}{l^2} & -\frac{3.\,3321}l
\end{array}
\right)
\]
has the eigenvalues
\[
-1.666/l-5.787i/l,-1.666/l+5.787i/l.
\]
This is a stable critical point, where
\[
h=-1.\,613/l,f=0.
\]

In the{\bf \ }second case, the functions $H$, $h$ and $f$ satisfy the
equations (33-36) which now read

\begin{equation}
f^2=\frac 4{1-8\gamma }H^2+\frac 8{1-8\gamma }Hh+\frac{4\left( 6\gamma
+1\right) }{1-8\gamma }h^2-\frac{16}{\left( 1-8\gamma \right) l^2},
\end{equation}

\begin{equation}
-2\stackrel{\cdot }{H}+2\left( 2\gamma -1\right) \stackrel{\cdot }{h}%
-3H^2+4\left( 4\gamma -1\right) Hh+\left( 2\gamma -1\right) h^2+\frac{%
8\gamma +1}4f^2+12l^{-2}=0,
\end{equation}
\begin{equation}
\stackrel{\cdot \cdot }{H}+\stackrel{\cdot \cdot }{h}+2\left( H+h\right)
\stackrel{\cdot }{H}+\left( H+h\right) \stackrel{\cdot }{h}+hH^2+h^2H+\frac 1%
{12\alpha }h=0,
\end{equation}
\begin{equation}
\allowbreak \stackrel{\cdot }{H}+\stackrel{\cdot }{h}+H^2+Hh-\frac{2\gamma }{%
\allowbreak 3\alpha }+\frac 1{12\alpha }=0.
\end{equation}
Eqs. (67), (68) and (70) give
\[
\stackrel{\cdot }{h}=\frac 4{8\gamma -1}H^2-\allowbreak \frac{4\left(
8\gamma -3\right) }{8\gamma -1}Hh+\frac{2\left( 4\gamma +3\right) }{8\gamma
-1}h^2-\frac{2\left( 16\gamma -1\right) }{\gamma \left( 8\gamma -1\right) l^2%
}+\frac{8\gamma -1}{24\gamma \alpha }=0.
\]
\[
\stackrel{\cdot }{H}=-\frac{8\gamma +3}{8\gamma -1}H^2+\frac{24\gamma -11}{%
8\gamma -1}\allowbreak Hh-2\frac{4\gamma +3}{8\gamma -1}h^2+\allowbreak 2%
\frac{16\gamma -1}{\gamma \left( 8\gamma -1\right) l^2}+\frac{\left( 8\gamma
-1\right) \left( 2\gamma -1\right) }{24\gamma \alpha },
\]
and then
\[
\stackrel{\cdot \cdot }{h}{}=\frac 8{8\gamma -1}H\stackrel{\cdot }{H}-4\frac{%
-3+8\gamma }{8\gamma -1}h\stackrel{\cdot }{H}-4\frac{-3+8\gamma }{8\gamma -1}%
H\stackrel{\cdot }{h}+\frac{4\left( 3+4\gamma \right) }{\left( -1+8\gamma
\right) }h\stackrel{\cdot }{h},
\]
\[
\stackrel{\cdot \cdot }{H}=-2\frac{3+8\gamma }{8\gamma -1}H\stackrel{\cdot }{%
H}+\frac{24\gamma -11}{8\gamma -1}h\stackrel{\cdot }{H}+\frac{24\gamma -11}{%
8\gamma -1}\allowbreak H\stackrel{\cdot }{h}-\frac{4\left( 3+4\gamma \right)
}{\left( -1+8\gamma \right) }h\stackrel{\cdot }{h}.
\]
Substituting into (69) yields
\[
h\left( \stackrel{\cdot }{H}+\stackrel{\cdot }{h}\right) +hH^2+h^2H+\frac 1{%
12\alpha }h=0.
\]
This equation and (70) lead to
\[
h=0.
\]
Then (69) becomes
\[
\stackrel{\cdot \cdot }{H}+2H\stackrel{\cdot }{H}=0.
\]
It has the solution
\[
\stackrel{\cdot }{H}=-H^2+C,
\]
\[
H=\sqrt{C}\frac{e^{2\sqrt{C}\left( t-t_0\right) }+1}{e^{2\sqrt{C}\left(
t-t_0\right) }-1}.
\]
The deceleration parameter is
\[
q=-\frac{\stackrel{\cdot }{H}}{H^2}-1=-\frac C{H^2}.
\]

When
\[
C=0
\]
we have
\[
-\frac{dH}{H^2}=dt,
\]
and then
\[
\frac 1H-\frac 1{H_0}=t-t_0.
\]

\subsection{When $\beta =-4\alpha $}

In this case the gravitational Lagrangian{\em \ }is the square of the
traceless Ricci tensor $\widetilde{R}_{\mu \nu }=R_{\mu \nu }-\frac 14g_{\mu
\nu }R$ [10].

According to section III, the equation (37) gives two cases.

In the first case $f=0$, the nonvanishing functions $H$ and $h$ satisfy the
equations (40), (41) and (44), i.e.,

\begin{eqnarray}
&&\left( \stackrel{\cdot }{H}+\stackrel{\cdot }{h}\right) ^2+2\left(
\stackrel{\cdot }{H}+\stackrel{\cdot }{h}\right) H\left( H+h\right) -h\left(
h+2H\right) \left( h+H\right) ^2  \nonumber \\
&&+\frac{6\gamma +1}{4\alpha }h^2+\frac 1{4\alpha }H^2+\frac 1{2\alpha }Hh-%
\frac 1\alpha l^{-2}=0,
\end{eqnarray}
\begin{equation}
4\left( \stackrel{\cdot }{H}+\stackrel{\cdot }{h}\right) -8\gamma \stackrel{%
\cdot }{h}+\allowbreak 4\left( 2\gamma +1\right) h^2-4\left( 8\gamma
-3\right) hH+8H^2-32l^{-2}=0,
\end{equation}
\begin{equation}
\left( \stackrel{\cdot \cdot }{H}+\stackrel{\cdot \cdot }{h}\right) -\left( H%
\stackrel{\cdot }{h}+h\stackrel{\cdot }{h}+h^2H\right) -h^3+\frac 1{8\alpha }%
h=0.
\end{equation}
They can be rewritten as\newline
\begin{eqnarray}
\stackrel{\cdot \cdot }{h} &=&\left( \frac{4\gamma +3}{2\gamma }h-\frac{%
8\gamma -4}{2\gamma }H\right) \stackrel{\cdot }{h}-\left( \frac{8\gamma -3}{%
2\gamma }h-\frac 2\gamma H\right) \stackrel{\cdot }{H}  \nonumber \\
&&+\frac 1{2\gamma }h^2H+\frac 1{2\gamma }h^3-\frac 1{16\alpha \gamma }h,
\end{eqnarray}
\begin{eqnarray}
\stackrel{\cdot \cdot }{H} &=&\left( -\frac{2\gamma +3}{2\gamma }h+\frac{%
5\gamma -2}\gamma H\right) \stackrel{\cdot }{h}+\left( \frac{8\gamma -3}{%
2\gamma }h-\frac 2\gamma H\right) \stackrel{\cdot }{H}  \nonumber \\
&&+\frac{2\gamma -1}{2\gamma }h^2H+\frac{2\gamma -1}{2\gamma }h^3-\frac{%
2\gamma -1}{16\alpha \gamma }h,
\end{eqnarray}
and
\begin{eqnarray}
&&4\gamma ^2\stackrel{\cdot }{h}^2-4\gamma \left( \left( 2\gamma +1\right)
h^2-\allowbreak \allowbreak \left( 8\gamma -2\right) Hh\allowbreak +H^2-%
\frac 8{l^2}\right) \stackrel{\cdot }{h}  \nonumber \\
&&+\allowbreak 4\gamma \left( 1+\gamma \right) h^4\allowbreak -8\gamma
\left( 1+4\gamma \right) h^3H+\allowbreak 4\gamma \left( 16\gamma -7\right)
h^2H^2\allowbreak -16\gamma H^3h  \nonumber \\
&&+\left( -\frac{16\left( 2\gamma +1\right) }{l^2}+\frac{6\gamma +1}{4\alpha
}\right) h^2+\left( \allowbreak \allowbreak 32\frac{4\gamma -1}{l^2}+\frac 1{%
2\alpha }\right) Hh  \nonumber \\
&&-32\frac{H^2}{l^2}+\left( -\frac{16}{l^2}+\frac 1{4\alpha }\right) H^2+%
\frac{64}{l^4}-\frac 1\alpha l^{-2}  \nonumber \\
&=&0.
\end{eqnarray}
Let
\[
\stackrel{\cdot }{H}=X,\stackrel{\cdot }{h}=Y.
\]
We have
\begin{eqnarray}
\stackrel{\cdot }{Y} &=&\left( \frac{4\gamma +3}{2\gamma }h-\frac{8\gamma -4%
}{2\gamma }H\right) Y-\left( \frac{8\gamma -3}{2\gamma }h-\frac 2\gamma
H\right) X  \nonumber \\
&&+\frac 1{2\gamma }h^2H+\frac 1{2\gamma }h^3-\frac 1{16\alpha \gamma }h, \\
\stackrel{\cdot }{X} &=&\left( -\frac{2\gamma +3}{2\gamma }h+\frac{5\gamma -2%
}\gamma H\right) Y+\left( \frac{8\gamma -3}{2\gamma }h-\frac 2\gamma
H\right) X  \nonumber \\
&&+\frac{2\gamma -1}{2\gamma }h^2H+\frac{2\gamma -1}{2\gamma }h^3-\frac{%
2\gamma -1}{16\alpha \gamma }h,
\end{eqnarray}
and
\begin{eqnarray}
&&\allowbreak Y^2-\left( \frac{\left( 2\gamma +1\right) }\gamma h^2-\frac{%
2\left( 4\gamma -1\right) }\gamma hH+\frac 1\gamma H^2-\frac 8{\gamma l^2}%
\right) Y  \nonumber \\
&&+\frac{\left( \gamma +1\right) }\gamma h^4\allowbreak -\frac{2\left(
4\gamma +1\right) }\gamma h^3H+\allowbreak \frac{\left( 16\gamma -7\right) }%
\gamma h^2H^2-\allowbreak \frac 4\gamma hH^3  \nonumber \\
&&+\left( -\frac{4\left( 2\gamma +1\right) }{\gamma ^2l^2}\allowbreak +\frac{%
6\gamma +1}{16\alpha \gamma ^2}\right) h^2+\left( \frac{8\left( 4\gamma
-1\right) }{\gamma ^2l^2}+\frac 1{8\alpha \gamma ^2}\right) Hh  \nonumber \\
&&+\left( -\frac 4{\gamma ^2l^2}+\frac 1{16\alpha \gamma ^2}\right) H^2+%
\frac{16}{\gamma ^2l^4}-\frac 1{4\alpha \gamma ^2}l^{-2}  \nonumber \\
&=&0.
\end{eqnarray}
The constraint equation (79) has the roots
\begin{equation}
Y=-\frac b2\pm \sqrt{\left( \frac b2\right) ^2-c}=Y\left( H,h\right) ,
\end{equation}
where

\begin{eqnarray}
b &=&-\left( \frac{\left( 2\gamma +1\right) }\gamma h^2-\frac{2\left(
4\gamma -1\right) }\gamma hH+\frac 1\gamma H^2-\frac 8{\gamma l^2}\right) \\
c &=&\frac{\left( \gamma +1\right) }\gamma h^4\allowbreak -\frac{2\left(
4\gamma +1\right) }\gamma h^3H+\allowbreak \frac{\left( 16\gamma -7\right) }%
\gamma h^2H^2-\allowbreak \frac 4{\gamma ^2}hH^3\gamma  \nonumber \\
&&+\left( -\frac{4\left( 2\gamma +1\right) }{\gamma ^2l^2}\allowbreak +\frac{%
6\gamma +1}{16\alpha \gamma ^2}\right) h^2+\left( \frac{8\left( 4\gamma
-1\right) }{\gamma ^2l^2}+\frac 1{8\alpha \gamma ^2}\right) Hh  \nonumber \\
&&+\left( -\frac 4{\gamma ^2l^2}+\frac 1{16\alpha \gamma ^2}\right) H^2+%
\frac{16}{\gamma ^2l^4}-\frac 1{4\alpha \gamma ^2}l^{-2},
\end{eqnarray}

So we are left with only three independent unknown functions $h$, $H$, and $%
X $, which satisfies the equations
\begin{eqnarray}
\stackrel{\cdot }{H} &=&X,  \nonumber \\
\stackrel{\cdot }{h} &=&Y\left( H,h\right) ,  \nonumber \\
\stackrel{\cdot }{X} &=&\left( -\frac{2\gamma +3}{2\gamma }h+\frac{5\gamma -2%
}\gamma H\right) Y\left( H,h\right) +\left( \frac{8\gamma -3}{2\gamma }h-%
\frac 2\gamma H\right) X  \nonumber \\
&&+\frac{2\gamma -1}{2\gamma }h^2H+\frac{2\gamma -1}{2\gamma }h^3-\frac{%
2\gamma -1}{16\alpha \gamma }h.
\end{eqnarray}

The critical point equations are
\begin{eqnarray}
X &=&0, \\
Y\left( H,h\right) &=&0, \\
\frac{2\gamma -1}{2\gamma }h^2H+\frac{2\gamma -1}{2\gamma }h^3-\frac{2\gamma
-1}{16\alpha \gamma }h &=&0.
\end{eqnarray}
The Eq.(86) means
\begin{equation}
h=0,
\end{equation}
or
\begin{equation}
hH+h^2-\frac 1{8\alpha }=0.
\end{equation}

For
\[
h=0,
\]
the equations (85) has the roots
\[
H=\pm \left( \frac 2l\right) .
\]
$\allowbreak $So we have the first pair of critical points
\[
X=0,h=0,H=\pm \left( \frac 2l\right) .
\]
For
\[
hH+h^2-\frac 1{8\alpha }=0,
\]
the critical point equations become

\begin{eqnarray}
X &=&0, \\
H &=&\left( -h+\frac 1{8\alpha h}\right) , \\
&&\allowbreak h^6-\left( \frac 1{5\alpha }-\frac 3{200\gamma \alpha }+\frac 8%
{5\gamma l^2}\right) h^4  \nonumber \\
&&+\left( \frac 1{100\alpha ^2}+\allowbreak \frac 1{320\gamma \alpha ^2}+%
\frac{16\gamma -1}{100\gamma ^2l^2\alpha }+\frac{16}{25\gamma ^2l^4}\right)
h^2  \nonumber \\
&&-\frac{8\gamma -1}{25600\alpha ^3\gamma ^2}\allowbreak -\frac 1{400\gamma
^2l^2\alpha ^2}  \nonumber \\
&=&0.
\end{eqnarray}
The equation (91) has the roots
\begin{eqnarray}
h_1^2 &=&\left( -\frac q2+\sqrt{\Delta }\right) ^{1/3}+\left( -\frac q2-%
\sqrt{\Delta }\right) ^{1/3}-\frac A3,  \nonumber \\
h_2^2 &=&\left( -\frac q2+\sqrt{\Delta }\right) ^{1/3}\omega +\left( -\frac q%
2-\sqrt{\Delta }\right) ^{1/3}\omega ^2-\frac A3,  \nonumber \\
h_3^2 &=&\left( -\frac q2+\sqrt{\Delta }\right) ^{1/3}\omega ^2+\left( -%
\frac q2-\sqrt{\Delta }\right) ^{1/3}\omega -\frac A3,
\end{eqnarray}
where
\begin{eqnarray}
\Delta &=&\left( \frac q2\right) ^2+\left( \frac p3\right) ^3,  \nonumber \\
\omega &=&\frac 12\left( -1+\sqrt{3}i\right) ,  \nonumber \\
p &=&B-\frac 13A^2,  \nonumber \\
q &=&\frac 2{27}A^3-\frac 13AB+C,
\end{eqnarray}
with
\begin{eqnarray}
A &=&-\left( \frac 1{5\alpha }-\frac 3{200\gamma \alpha }+\frac 8{5\gamma l^2%
}\right) ,-\frac A3=+\frac 1{15\alpha }-\frac 1{200\gamma \alpha }+\frac 8{%
15\gamma l^2}  \nonumber \\
B &=&\frac 1{100\alpha ^2}+\allowbreak \frac 1{320\gamma \alpha ^2}+\frac{%
16\gamma -1}{100\gamma ^2l^2\alpha }+\frac{16}{25\gamma ^2l^4}  \nonumber \\
C &=&-\left( \frac{8\gamma -1}{25600\alpha ^3\gamma ^2}\allowbreak +\frac 1{%
400\gamma ^2l^2\alpha ^2}\right) .
\end{eqnarray}
The equations (89), (90) and (92) give the critical points $\left\{
X,H,h\right\} $. Every one of these point corresponds to a de Sitter
spacetime.

The dynamical system (83) has the Jacobian elements
\begin{eqnarray}
\frac{\partial \stackrel{\cdot }{H}}{\partial H} &=&0,\frac{\partial
\stackrel{\cdot }{H}}{\partial h}=0,\frac{\partial \stackrel{\cdot }{H}}{%
\partial X}=1,  \nonumber \\
\frac{\partial \stackrel{\cdot }{h}}{\partial H} &=&\frac{\partial Y}{%
\partial H},\frac{\partial \stackrel{\cdot }{h}}{\partial h}=\frac{\partial Y%
}{\partial h},\frac{\partial \stackrel{\cdot }{h}}{\partial X}=0,  \nonumber
\\
\frac{\partial \stackrel{\cdot }{X}}{\partial H} &=&\frac{5\gamma -2}\gamma
Y\left( H,h\right) +\left( -\frac{2\gamma +3}{2\gamma }h+\frac{5\gamma -2}%
\gamma H\right) \frac{\partial Y}{\partial H}-\frac 2\gamma X+\frac{2\gamma
-1}{2\gamma }h^2,  \nonumber \\
\frac{\partial \stackrel{\cdot }{X}}{\partial h} &=&-\frac{2\gamma +3}{%
2\gamma }Y\left( H,h\right) +\left( -\frac{2\gamma +3}{2\gamma }h+\frac{%
5\gamma -2}\gamma H\right) \frac{\partial Y}{\partial h}+\frac{8\gamma -3}{%
2\gamma }X  \nonumber \\
&&+\frac{2\gamma -1}\gamma hH+3\frac{2\gamma -1}{2\gamma }h^2-\frac{2\gamma
-1}{16\alpha \gamma },  \nonumber \\
\frac{\partial \stackrel{\cdot }{X}}{\partial X} &=&\frac{8\gamma -3}{%
2\gamma }h-\frac 2\gamma H,
\end{eqnarray}
where
\begin{eqnarray}
\frac{\partial Y}{\partial H} &=&-\frac{\left( 4\gamma -1\right) }\gamma h+%
\frac 1\gamma H  \nonumber \\
&&\pm \frac 1{2\sqrt{\left( \frac b2\right) ^2-c}}\left( \frac 1{\gamma ^2}%
h^3+\allowbreak \allowbreak \frac 3{\gamma ^2}h^2H+\allowbreak \frac 3{%
\gamma ^2}hH^2+\frac 1{\gamma ^2}H^3-\frac 1{8\gamma ^2\alpha }h-\frac 1{%
8\alpha \gamma ^2}H\right) ,
\end{eqnarray}
\begin{eqnarray}
\frac{\partial Y}{\partial h} &=&\frac{\left( 2\gamma +1\right) }\gamma h-%
\frac{\left( 4\gamma -1\right) }\gamma H  \nonumber \\
&&\pm \frac 1{2\sqrt{\left( \frac b2\right) ^2-c}}\left( \frac 1{\gamma ^2}%
h^3+\allowbreak \frac 3{\gamma ^2}h^2H+\allowbreak \allowbreak \frac 3{%
\gamma ^2}hH^2+\allowbreak \frac 1{\gamma ^2}H^3-\frac{6\gamma +1}{8\gamma
^2\alpha }h\allowbreak -\frac 1{8\gamma ^2\alpha }H\right) .
\end{eqnarray}

In order to analyze their stability we give the parameter $\alpha $ and $%
\gamma $ specific values and then obtain the results:

For the critical point $X=0,h=0,H=2/l$, corresponding calculation indicates
it is unstable for $\alpha =\frac 1{32}l^2.$

In the case $X=0,$ $hH+h^2-\frac 1{8\alpha }=0$, we have

When
\[
\alpha =\left( \frac 1{32}l^2\right) ,\gamma =\left( \frac 14\right)
\]
the equations (92) and (90) give
\begin{eqnarray*}
h_1^2 &=&\frac{1.\,9814}{l^2},h_1=\pm \frac{1.\,4076}l,H_1=\pm \frac{1.\,4341%
}l \\
h_2^2 &=&\frac{4.\,4493+3.\,3485i}{l^2}, \\
h_3^2 &=&\frac{4.\,4493-3.\,3485i}{l^2},
\end{eqnarray*}

At
\[
h_1=\left( \frac{1.\,4076}l\right) ,H_1=\left( \frac{1.\,4341}l\right) ,
\]
for
\[
Y=-\frac b2+\sqrt{\left( \frac b2\right) ^2-c}=0,
\]
the dynamical system (83) has the form
\begin{eqnarray*}
\stackrel{\cdot }{H} &=&X, \\
\stackrel{\cdot }{h} &=&Y\left( H,h\right) , \\
\stackrel{\cdot }{X} &=&-\left( 7h+3H\right) Y\left( H,h\right) -\left(
7h+3H\right) X \\
&&-h^2H-h^3+\frac 4{l^2}h,
\end{eqnarray*}
with
\begin{eqnarray*}
Y &=&3h^2+2H^2-\frac{16}{l^2} \\
&&+(4h^4+24h^2H^2-\frac{80}{l^2}h^2+4H^4-\frac{32}{l^2}H^2 \\
&&+\frac{128}{l^4}+16h^3H+\allowbreak 16hH^3-64h\frac H{l^2})^{1/2}.
\end{eqnarray*}
Its Jacobian
\[
{\cal M}=\left(
\begin{array}{lll}
0 & 0 & 1 \\
\frac{21.\,324}l & \frac{12.\,665}l & 0 \\
-\frac{303.\,83}{l^2} & -\frac{185.\,26}{l^2} & -\frac{14.\,288}l
\end{array}
\right)
\]
has the eigenvalues: $-0.\,39225/l+11.\,048i/l,\allowbreak
-0.\,39225/l-11.\,048i/l,\allowbreak -0.\,8385/l$. The critical point
\[
h_1=\left( \frac{1.\,4076}l\right) ,H_1=\left( \frac{1.\,4341}l\right) ,
\]
is stable, where $f=0$.

For
\[
Y=-\frac b2-\sqrt{\left( \frac b2\right) ^2-c}=-\frac{11.\,886}{l^2},
\]
the dynamical system (83) has the Jacobian

\[
{\cal M}=\left(
\begin{array}{lll}
0 & 0 & 1 \\
-\frac{9.\,8509}l & \frac{4.\,2259}l & 0 \\
\frac{173.\,12}{l^2} & \frac{17.\,401}{l^2} & -\frac{14.\,288}l
\end{array}
\right)
\]
with the eigenvalues: $-22.\,33/l,\allowbreak
6.\,1339/l+1.\,6776i/l,\allowbreak 6.\,1339/l-1.\,6776i/l$. The critical
point is unstable.

In the case $f\neq 0$, (33), (34), (35) and (39) read (in vacuum)

\begin{eqnarray}
&&\left( \stackrel{\cdot }{H}+\stackrel{\cdot }{h}\right) ^2+2\left(
\stackrel{\cdot }{H}+\stackrel{\cdot }{h}\right) H\left( H+h\right)
\nonumber \\
&&-h\left( h+2H\right) \left( h+H\right) ^2+\frac 12\left( h+H\right)
^2f^2-\allowbreak \frac 1{16}f^4  \nonumber \\
&&+\frac{6\gamma +1}{4\alpha }h^2+\frac 1{4\alpha }H^2+\frac 1{2\alpha }Hh+%
\frac{8\gamma -1}{16\alpha }f^2-\frac 1\alpha l^{-2}=0,
\end{eqnarray}

\begin{equation}
\left( \stackrel{\cdot \cdot }{H}+\stackrel{\cdot \cdot }{h}\right) -\left( H%
\stackrel{\cdot }{h}+h\stackrel{\cdot }{h}+h^2H\right) -\frac 12\left( 2h^3-f%
\stackrel{\cdot }{f}-\frac 12hf^2\right) +\frac 1{8\alpha }h=0,
\end{equation}
\begin{equation}
4\alpha \left( \stackrel{\cdot }{H}+\stackrel{\cdot }{h}\right) \allowbreak
-4\alpha Hh-\alpha \ \left( 4h^2-f^2\right) -4\gamma +\frac 12=0.
\end{equation}
\begin{equation}
f^2=4\left( \stackrel{\cdot }{H}+\stackrel{\cdot }{h}\right) -8\gamma
\stackrel{\cdot }{h}+\allowbreak 4\left( 2\gamma +1\right) h^2-4\left(
8\gamma -3\right) hH+8H^2-32l^{-2},
\end{equation}

These equations have the solution
\begin{eqnarray}
\alpha &=&\left( \frac 1{64}-\frac 14\gamma \right) l^2, \\
H^2 &=&\frac{32\gamma }{\left( 16\gamma -1\right) l^2}, \\
f^2 &=&\allowbreak -\frac{32\left( 8\gamma -1\right) }{l^2\left( 16\gamma
-1\right) }.
\end{eqnarray}
For
\[
\gamma >\frac 1{16},
\]
or
\[
\gamma <0,
\]
we have a de Sitter solution
\begin{equation}
H=\frac 4l\sqrt{\frac{2\gamma }{16\gamma -1}}.
\end{equation}
When
\[
\left| \gamma \right| \gg 1,
\]
we have
\[
H^2\approx \frac 2{l^2},
\]
a value speculated [14]. When
\[
\frac 1{16}<\gamma \leq \frac 18
\]
$f$ is real. In this case, it is the pseudotrace axial ingredient $f$ of
torsion that produces the effect of acceleration of cosmological expansion.

\section{Conclusions}

Stating from a de Sitter gauge theory a gravitational Lagrangian (13) which
is identified with the Lagrangian of quadratic-curvature gravities with
torsion has been constructed. The cosmological equations (33-36) for spatial
flat universe have been obtained. To search for vacuum solutions of them in
two specific models, the conformal model and the zero-energy (Deser-Tekin)
model, the dynamical systems have been derived, some de Sitter critical
points and their stability have been investigated. These points are always
exact constant solutions in the context of autonomous dynamical systems and
describe the asymptotic behavior. Some stable de Sitter critical points have
been found. For any physical theories, to find exact mathematical solutions
is an important topic. Next comes the physical interpretations of the
solution thus obtained. Mathematically,\ de Sitter as the maximally space is
undoubtedly important for any gravity theories. From observational side,
recent studies illuminate that both the early universe (inflation) and the
late-time universe (cosmic acceleration) can be regarded as fluctuations on
a de Sitter background. So de Sitter takes a pivotal status in gravity,
especially in modern cosmology.

The solutions in section IV indicate that when $f=0$, $h\neq 0$, the
cosmological equations have stable de Sitter critical points. This means
that the scalar ingredient $h$ of{\em \ }torsion could be considered as a
''phantom'' field, since it does not interact directly with matter; it only
interacts indirectly via gravitation. In the case $f\neq 0$, $h=0$, it is
the pseudotrace axial ingredient $f$ of torsion that produces the effect of
acceleration of cosmological expansion. Therefore the spacetime in the
vacuum has the structure of de Sitter spacetime with torsion including the
pseudotrace axial ingredient $f$ as well as the scalar ingredient $h$.

In summary, in the framework of gauge theory of gravity some cosmological
models can be constructed to explain observable acceleration of cosmological
expansion. The effect of acceleration of cosmological expansion in these
models has the geometrical nature and is connected with geometrical
structure of physical spacetime. The spacetime in the vacuum has the
structure of de Sitter spacetime with torsion.

\end{document}